\documentclass[pra,twocolumn,groupedaddress]{revtex4-2}
\usepackage{graphicx}
\usepackage{amsmath}
\usepackage{amssymb}
\usepackage{bm}
\usepackage{color}
\usepackage{hyperref}
\usepackage{tabularx}
\usepackage{multirow}
\usepackage{braket}
\usepackage{epstopdf}
\usepackage{lineno}

\begin{document}

\title{
Spin-textured Chern bands in AB-stacked transition metal dichalcogenide bilayers}

\author{Yang Zhang}
\email{yzhang11@mit.edu}
\author{Trithep Devakul}
\author{Liang Fu}
\email{liangfu@mit.edu}
\affiliation{Department of Physics, Massachusetts Institute of Technology, Cambridge, Massachusetts 02139, USA}

\begin{abstract}
While transition metal dichalcogenide (TMD) based moir\'e materials have been shown to host various correlated electronic phenomena, topological states have not been experimentally observed until now~\cite{li2021quantum}.
In this work, using first principle calculations and continuum modeling, we reveal the displacement field induced topological moir\'e bands in AB-stacked TMD heterobilayer MoTe$_2$/WSe$_2$.
Valley contrasting Chern bands with non-trivial spin texture are formed from interlayer hybridization between MoTe$_2$ and WSe$_2$ bands of nominally opposite spins.
Our study establishes a recipe for creating topological bands in AB stacked TMD bilayers in general, which provides a highly tunable platform for realizing quantum spin Hall and interaction induced quantum anomalous Hall effects.
\end{abstract}

\maketitle

Two-dimensional (2D) transition metal dichalcogenides (TMD) MX$_2$ exhibit a host of interesting
electronic phases, such as charge density wave and Ising superconductivity 
\cite{xi2016ising,sohn2018unusual} and magnetism 
\cite{bonilla2018strong}, ferroelectricity 
\cite{yuan2019room,wang2021ferroelectricity} 
and quantum spin Hall insulator 
\cite{qian2014quantum,wu2018observation,tang2017quantum,shi2019imaging,sajadi2018gate}.
More recently, semiconductor TMD moir\'e superlattices \cite{Tang2020,Regan2020,Shabani2021,wang2020correlated,li2021imaging,Wu2018,Zhang2020, shimazaki2020strongly,PhysRevX.11.021027,Jin2021,li2021continuous,Wu2019,Zhang2021,Slagle2020,devakul2021magic,padhi2021generalized,Xu2020,Bi2021,Pan2020,MoralesDuran2020,Zang2021,padhi2021generalized,Cazalilla2014}
have emerged as a highly tunable
quantum simulator of correlated electron physics \cite{kennes2021moire}. Unlike graphene with highly mobile Dirac electrons,
semiconductor TMDs have an intrinsic large effective mass that strongly enhances interaction effects at low density.

In TMD heterobilayer WSe$_2$/WS$_2$ \cite{Tang2020,Regan2020,Shabani2021,li2021imaging,Wu2018, Zhang2020}, the moir\'e pattern due to lattice mismatch
 introduces a periodic effective potential for low-energy electrons or holes.
At long moir\'e wavelength, interaction and potential energies dominate over the kinetic energy,  
leading to a plethora of strong-coupling phases including Mott insulators,
generalized Wigner crystals and stripe charge order.
In twisted homobilayer WSe$_2$, the bandwidth is continuously tunable by the twist angle $\theta$.
At $\theta \sim 4^\circ$--$5^\circ$ where the bandwidth
is comparable to interaction energy, metal-insulator transitions and quantum critical phenomena
tuned by the displacement field have been observed at half band filling \cite{wang2020correlated}.
A candidate state for the correlated insulator is an intervalley spin density wave \cite{Bi2021}
that is smoothly connected to an $xy$-ordered antiferromagnet \cite{Pan2020,MoralesDuran2020}.
Signs of superconductivity have also been observed below half filling \cite{wang2020correlated}. 
The remarkably rich physics of TMD moir\'e systems is being uncovered at a rapid pace.

Due to strong spin-orbit coupling and broken inversion symmetry,  monolayer TMDs in the 2H phase, such as WSe$_2$ and MoTe$_2$, feature large spin-orbit splitting in the valence band \cite{xiao2012coupled}.
The topmost valence bands at $\pm K$ are spin-polarized with $s_z=\uparrow, \downarrow$ respectively.
The spin-valley locking opens the possibility of topological bands and quantum spin Hall effect in TMD moir\'e systems \cite{cazalilla2014quantum}.
Using a continuum model approach, Wu {\it et al} predicted topologically nontrivial $\pm K$-valley moir\'e bands
with spin/valley Chern number in twisted homobilayer TMD \cite{Wu2019}.
Based on first-principles calculations,
we recently identified flat topological bands
in twisted bilayer WSe$_2$ and further predicted interaction-induced spin/valley polarization at half band filling,
which leads to a quantum anomalous Hall (QAH) insulator  \cite{devakul2021magic}.

The above theoretical works are concerned with TMD homobilayers with a small twist angle starting from the AA stacking, where every metal (M) or chalcogen (X) atom on the top layer is aligned with the same type of atom on the bottom layer.
The formation of topological bands relies on spin- and valley-conserving tunneling between the two layers.
In contrast, in AB-stacked TMD bilayers where the top layer is rotated by $180^\circ$ relative to the bottom layer,
the valence bands of two layers in the same valley carry different spins.
It is commonly believed that the spin mismatch blocks
interlayer tunneling and therefore the two layers are electronically decoupled.

Unexpectedly, a recent experiment
discovered a QAH state in AB-stacked MoTe$_2$/WSe$_2$ heterobilayer in a range of displacement fields \cite{li2021quantum}.
Magnetic hysteresis behavior is observed in both and Hall resistance and Kerr rotation below $5$K.
At zero field, the Hall resistance reaches the quantized value $h/e^2$ below $2.5$K.
The origin of this QAH state calls for theoretical understanding.

In this work, by performing large-scale first-principles calculations
we find that displacement field induces band inversion in AB-stacked MoTe$_2$/WSe$_2$ and leads to
valley Chern bands with nontrivial spin and layer characters. In contrast to twisted homobilayers,
interlayer coupling between  MoTe$_2$ and WSe$_2$ bands of {\it different spins}
is essential for gap opening after band inversion.
We introduce a new continuum model to describe the low energy moir\'e bands of TMD heterobilayers and
the displacement field induced topological transition.
Our findings suggest spontaneous valley polarization in topological bands as the mechanism for the observed QAH state in MoTe$_2$/WSe$_2$.

\begin{figure}[t]
\includegraphics[width=\columnwidth]{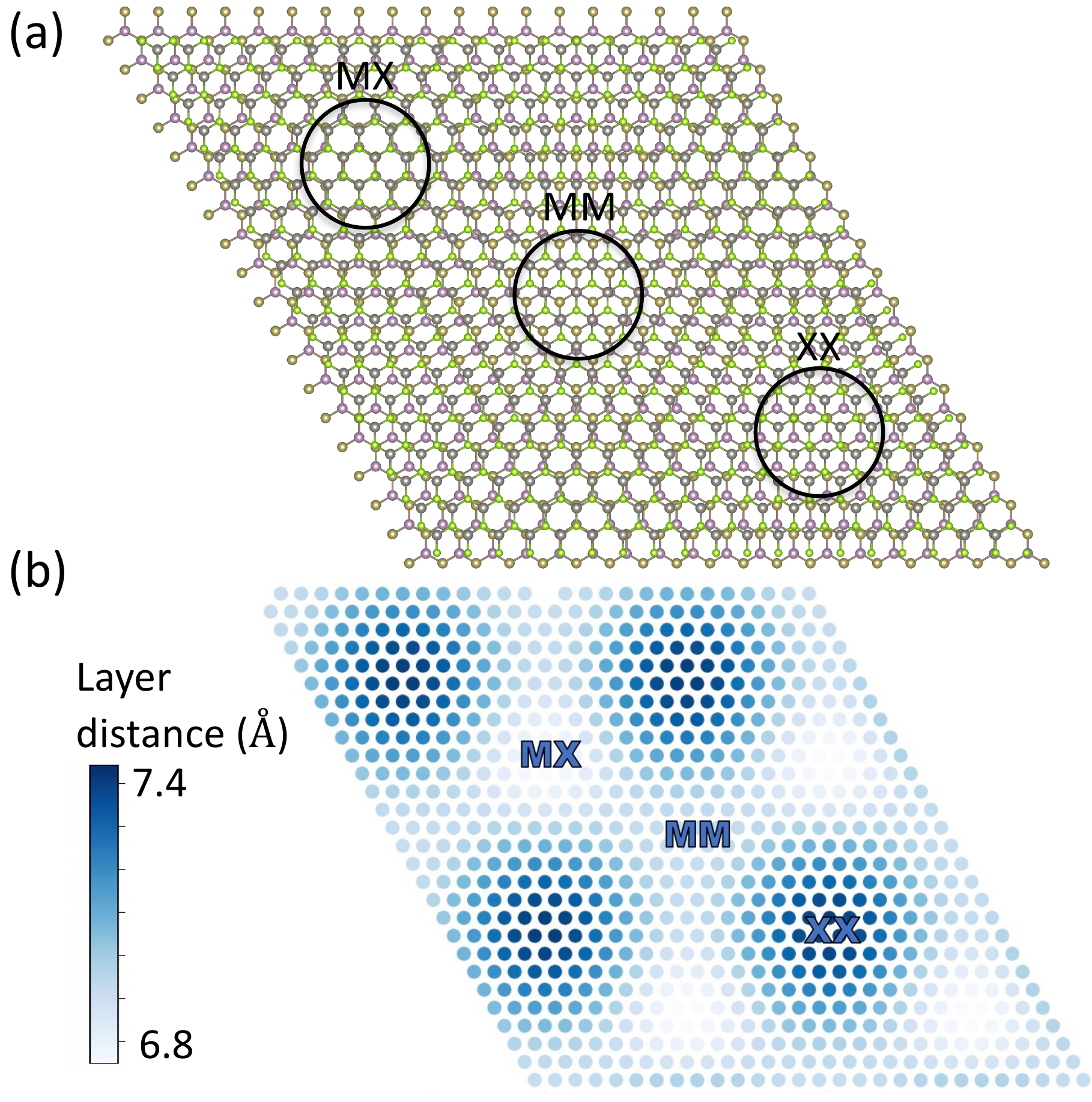}
\caption{
Lattice structure for AB stacked MoTe$_2$/WSe$_2$ heterobilayer.
(a) MM, XX and MX stacking regions; (b) spatial dependent layer distance in $2\times2$ moir\'e superlattice with strong out of plane corrugation.
}\label{fig1}
\end{figure}

We consider a heterobilayer TMD MoTe$_2$/WSe$_2$, with $a_b(a_t)$ as the lattice constant of bottom (top) layer, and $\theta$ as the twist angle. The lattice mismatch leads to a moir\'e superlattice in Fig. \ref{fig1}, with superlattice constant $a_{ M}=a_b/\sqrt{\delta^2+\theta^2}$ where $\delta=(a_b-a_t)/a_t$.
In AB stacked TMD bilayers, the orientation of top layer is anti-parallel to that of bottom layer.
Here, we study the moir\'e superlattice at $\theta=0$, where the moir\'e wavelength $a_M \sim 5$ nm is maximum.
In this superlattice, there are three high symmetry regions denoted as MM, XX, and MX.
In the MX region, the M atom on the bottom layer is locally aligned with the X atom on the top layer, and likewise for MM and XX, see Fig. \ref{fig1}a.
The bilayer structure in these stacking configurations is invariant under three-fold rotation about the $z$ axis, $C_{3z}$, centered at the high symmetry regions.

A unit cell of MoTe$_2$/WSe$_2$ superlattice at $\theta=0$ contains $13\times13$ MoTe$_2$ unit cells and $14\times14$ WSe$_2$ unit cells.
As well studied in previous works on various homobilayers and AA stacked heterobilayers \cite{li2021imaging}, lattice relaxation significantly modifies the low energy electronic states, especially the real space features of the moir\'e bands and the relative energy level of the $\Gamma$ and $K$ valley moir\'e bands \cite{Zhang2021,xian2020realization}.
In order to obtain accurate moir\'e lattice structures, we perform large-scale density functional theory calculations with the SCAN+rVV10 van der Waals density functional~\cite{peng2016versatile}, which captures the intermediate-range van der Waals interaction through its semilocal exchange term resulting in a better estimation of layer spacing.
Owing to the different local atomic alignments in MM, XX and MX regions, the layer distances have a strong spatial variation as shown in Fig. \ref{fig1}b, which are 6.8\AA{}, 7.0\AA{} and 7.4\AA{} for MX, MM and XX regions respectively.

\begin{figure}[t]
\includegraphics[width=\columnwidth]{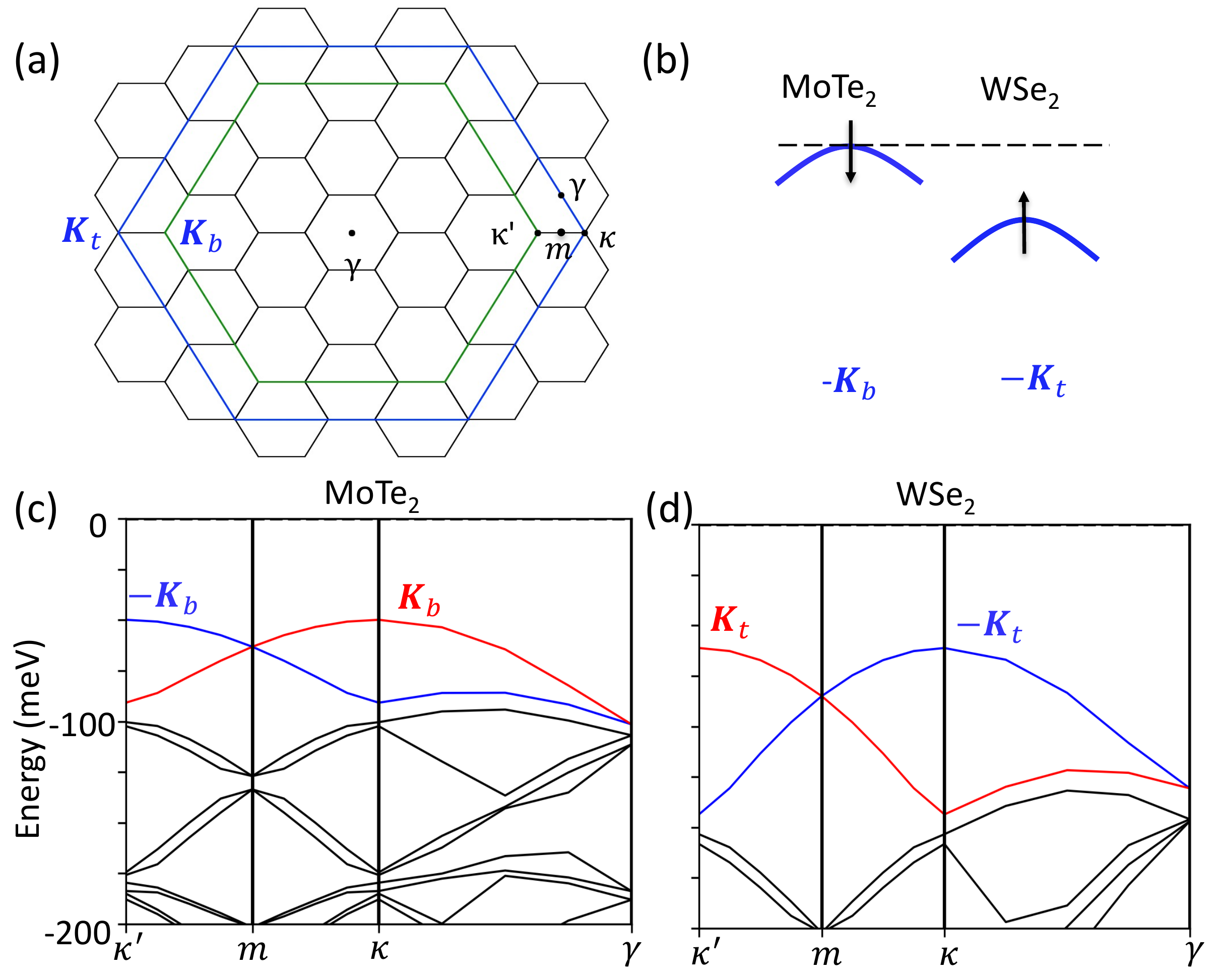}
\caption{
(a) Schematic figure for BZ folding. Black lines are mBZ of the moir\'e supercell, $\gamma$, $m$, $\kappa$, and $\kappa^\prime$ are labels in the mBZ. Green and blue lines are the full BZ for the primitive cell in the $3N+1$ and $3N+2$ supercell, which apply to 13$\times$13 MoTe$_2$ and $14\times 14$ WSe$_2$. (b) Band offset between MoTe$_2$ and WSe$_2$ $-K$ valley. DFT Band structures of (c) $13 \times 13$ MoTe$_2$ and (d) $14\times 14$ WSe$_2$ with lattice corrugation extracted from relaxed untwisted heterobilayers. Note we move the moir\'e bands in (c) and (d) to the same energy level, and band offset is therefore not presented.
}\label{fig:mono}
\end{figure}

At zero displacement field, there is a large valence band offset $\Delta E_g \sim 110$ meV (consistent with \cite{gong2013band})  between MoTe$_2$ and WSe$_2$ (Fig~\ref{fig:mono}(b)), and the holes are first filled to MoTe$_2$ layer.
The primary effect of the WSe$_2$ layer on MoTe$_2$, and vice versa, is through the lattice corrugation effect.
Therefore, we first study the moir\'e band structures for separated MoTe$_2$ layer and WSe$_2$ layer with corrugation extracted from relaxed aligned heterobilayers. The $z$ corrugation in WSe$_2$ layer is around $0.7$\AA{}. While the height variation in MoTe$_2$ is at the order of $0.2$\AA{}. Due to the relatively smaller moir\'e wavelength, here the built-in strain is 0.5\%, smaller than the 1\% in the case of 8.2 nm AA stacked WSe$_2$/WS$_2$ \cite{li2021imaging}.

Owing to strong Ising spin-orbit coupling in the valence band side, valley and spin are locked \citep{xiao2012coupled} in monolayer TMD, resulting in the top valence bands being mostly spin-$\uparrow$ at $+K$ valley and spin-$\downarrow$ at $-K$.
The moir\'e Brillouin zone (mBZ) is constructed by folding the full Brillouin (BZ) zone as illustrated in Fig. 2(a).
The $K$ and $-K$ valley fold to $\kappa$ and $\kappa^\prime$ respectively in the $13 \times 13$ MoTe$_2$ supercell, while in WSe$_2$ with $14\times 14$ supercell, the $K$ and $-K$ valley instead fold to $\kappa^\prime$ and $\kappa$.
Since the moir\'e wavelength is relatively small ($a_M \sim5$ nm), the kinetic energy scale $\sim -\frac{|\bm{\kappa}|^2}{2_{m_b}}\sim40$ meV ($m_b=0.65m_0$ is the effective mass of bottom layer MoTe$_2$, $m_t=0.35 m_0$ for top layer WSe$_2$ and therefore even larger kinetic energy) is much larger than the potential energy scale, giving rise to a nearly-free electron dispersion with small band gaps at the mini Brillouin zone boundaries in Fig. 2(c,d).

\begin{figure*}[ht]
\includegraphics[width=\textwidth]{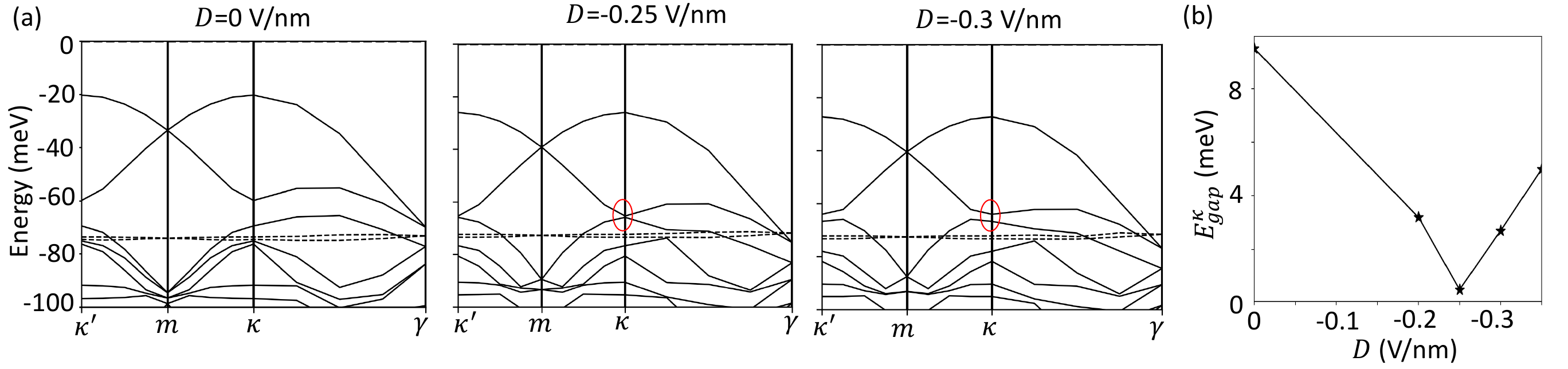}
\caption{
(a) DFT band structure of AB stacked MoTe$_2$/WSe$_2$ heterobilayer with displacement field D= 0, -0.25 (critical field) and -0.3 V/nm, showing gap reopening and band inversion. (b) Displacement field dependent band gap at $\kappa$, the band gap reduces from 9 meV to 0 meV and then increases to 4 meV.
}\label{fig:dft}
\end{figure*}

Due to the large valence band offset $\Delta E_g$ between MoTe$_2$ and WSe$_2$, we expect the top moir\'e valence bands in the heterobilayer to be quite similar to the MoTe$_2$ layer with corrugation.
When the band offset is reduced by displacement field, the top moir\'e bands of the WSe$_2$ layer is lifted and intersects with band bottom of first set of moir\'e bands in MoTe$_2$, causing a band inversion at $\kappa$ between the $-K$ valleys of MoTe$_2$ and $180^\circ$ rotated WSe$_2$, of mostly spin-$\downarrow$ and $\uparrow$ respectively.

To substantiate this picture, we calculate the band structure of AB stacked MoTe$_2$/WSe$_2$ heterobilayer
from density functional theory.
With zero out of plane displacement field, the top two valence moiré bands are well separated from the other moiré bands with a band gap around 10 meV at $\kappa$ in the mBZ.
Compared to the band structure of monolayer MoTe$_2$ with corrugation, it is clear that majority of the dispersive bands in heterobilayer MoTe$_2$/WSe$_2$ are developed from MoTe$_2$ layer, and in the energy range
of interest, there is only one additional moir\'e band coming from the WSe$_2$ band maximum as shown in Fig. 3(a).

When the displacement field strength is increased, the single particle band gap at $\kappa$ between the two $-K$ valley dispersive moiré bands gradually reduces from 9 meV and closes at a critical field of around $-0.25$ V/nm, as shown in Fig. 3(b). At a displacement field $-0.3$ V/nm, we observe a band inversion near $\kappa$ with a direct gap $\sim1.5$ meV.

Besides the dispersive moiré bands, we find two flat bands (dashed lines).
In order to determine the origin of various moiré bands, we plot the Kohn-Sham wavefunctions in real space.
The wavefunction of the flat band at $\kappa$ (dashed lines in Fig. 3(a)) is formed of the bonding and antibonding orbitals between MoTe$_2$ and WSe$_2$ and is therefore spread out in two layers, indicating that it arises from the $\Gamma$ valley of the monolayer TMDs.
The dispersive moiré valence bands at $\kappa$ are instead strongly localized at one layer with around 80\% weight from $d$ orbitals of metallic atoms, forming a triangular network, and we conclude that they arise from the $\pm K$ valley of MoTe$_2$ and WSe$_2$.
In real space, the first two $-K$ valley moiré bands are peaked at the MM region in MoTe$_2$ and XX region in WSe$_2$, respectively, forming an effective honeycomb lattice with band gap tunable by displacement field.

While the $\Gamma$-valley bands of monolayer WSe$_2$ and MoTe$_2$ are significantly lower in energy
than their $\pm K$-valley counterparts, interlayer tunneling in MoTe$_2$/WSe$_2$
substantially pushes up the energy of $\Gamma$-valley moir\'e bands,
which overlaps with the $\pm K$-valley moir\'e bands.
Nonetheless, due to large separation between $\Gamma$ and $\pm K$ in momentum space,
these two sets of moir\'e bands are decoupled at the single-particle level.
The $\Gamma$-valley flat bands are not relevant to our discussion concerning the band gap and
topological properties between two $\pm K$ valley moir\'e bands.

\begin{figure*}[ht]
\includegraphics[width=\textwidth]{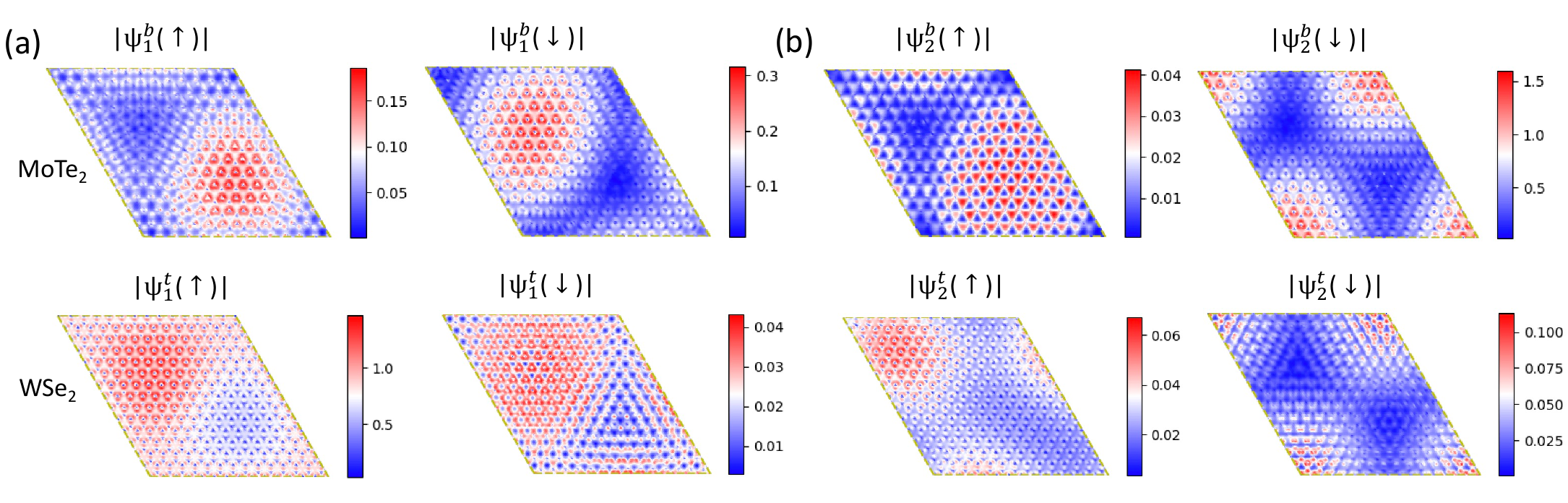}
\caption{
At $D=-0.3$ V/nm after topological transition, spin and layer resolved $\kappa$ wavefunction for (a) first topological band; (b) and second topological band. Both the spin and layer are mixed in these two bands. And the first topological band is dominantly localized at WSe$_2$ layer spin up channel, with sizeable component localized at MoTe$_2$ layer spin up channel.
}\label{fig:wave}
\end{figure*}

\begin{figure}[ht]
\includegraphics[width=0.4\textwidth]{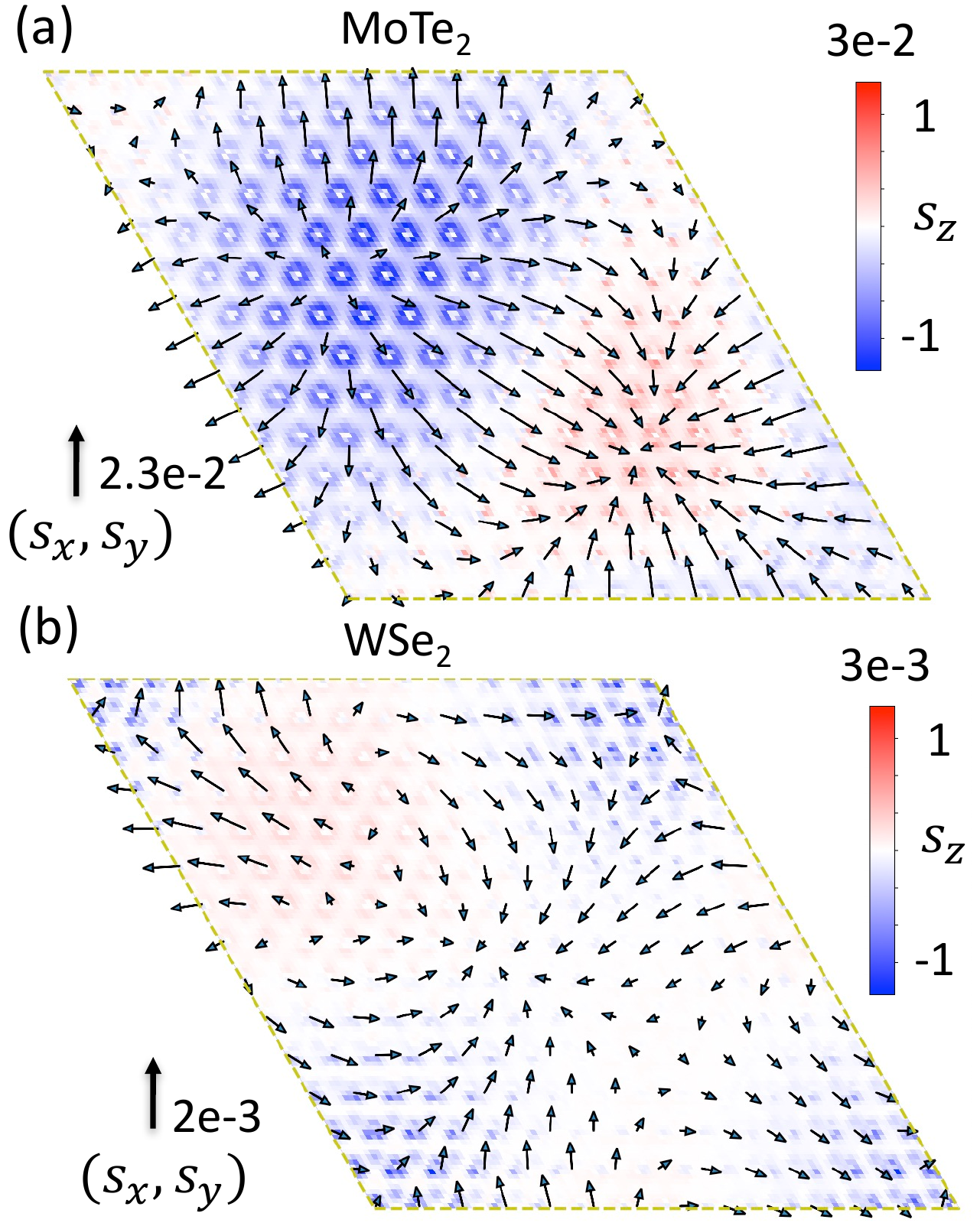}
\caption{
At $D=-0.3$ V/nm after topological transition, spin texture at $\kappa$ for (a) first topological band projected at MoTe$_2$ layer; (b) and second topological band projected at WSe$_2$ layer. We plot the $(s_x,s_y)$ as in plane vector and $s_z$ as colormap for both bands. The spin component in majority layer is almost fully up or down at $s_z$ direction. While the spin texture in minority layer forms a real space Skyrmion lattice.
}\label{fig:SKM}
\end{figure}

To reveal the origin of topological gap closing, we further analyze the wavefunction around
the band inversion point $\kappa$. Inspecting the band structure of separated monolayers, we find that the pair of topological moir\'e bands are developed from  $-K$ valley of MoTe$_2$ and $-K$ valley of MoTe$_2$, which should
have nominally opposite spins due to AB stacking.
Tunneling between these bands is only possible if spin $s_z$ is not fully conserved,
as is generally true in systems with strong spin-orbit coupling.

The validity of this scenario can be further tested by layer and spin resolved (projected in the $s_z$ direction) wavefunction of the two $-K$ valley bands at $\kappa$.
At the displacement field $D=-0.3$V/nm in the inverted regime,
the wavefunction of the first band is dominantly located at WSe$_2$ layer, with 6\% charge density at MoTe$_2$ layer shown in Fig. \ref{fig:wave}(a).
In the spin channel, the amplitude of spin up and spin down resolved wavefunctions in MoTe$_2$, $\sqrt{\int d\bm{r} |\psi_1^b(\uparrow)|^2/\int d\bm{r} |\psi_1^b(\downarrow)|^2}\sim 0.6$ is of comparable magnitude,
a direct proof of $s_z$ non-conservation and spin mixing in AB-stacked heterobilayer.
Even in corrugated monolayers, we find that $s_z$ is not conserved due to the breaking of mirror symmetry $m_z$ \cite{soriano2021spin}.

In Fig. \ref{fig:SKM}, we plot the layer resolved real space spin texture, $\bm{s}(\bm{r})$, for the two topological bands at $\kappa$.
For the top topological band, the majority layer WSe$_2$ is almost fully spin up.
The spin moment in the minority layer MoTe$_2$, on the other hand, very clearly forms a real space Skyrmion lattice,
with spins at the MX regions pointing down and spins at the XX regions pointing up, and a vortex structure in the in-plane spin around these regions.
In second topological band, the minority layer WSe$_2$ similarly exhibits a Skyrmion texture.

The absence of spin $s_z$ conservation allows interlayer tunneling between
same-valley bands in the AB-stacked heterobilayer.
We expect the tunneling amplitude to be much weaker in this case, compared the AA stacked case which involve tunneling between matching spin bands (which for WSe$_2$ homobilayer is around 18 meV~\cite{devakul2021magic}).
From the amplitude ratio between same spin wavefunctions in MoTe$_2$ and WSe$_2$, we estimate that the tunneling magnitude in our case should be of order $\sim 1$meV.
The relatively weak tunneling amplitude is fully consistent with
the small topological gap in the inverted regime,
which is $\sim$1.5 meV from DFT calculation.
This value is close to $\sim$2 meV inferred from capacitance measurement.
Thus, our DFT calculation and wavefunction analysis clearly show that there is a non-zero, albeit small, interlayer tunneling despite the two bands being of primarily opposite spin.


We now develop a continuum model for AB-stacked heterobilayers, which
 well captures the physics of the band inversion.
As a first step, the moire band structures of single layer MoTe$_2$ or WSe$_2$ can be modeled by
a continuum model with kinetic and potential term, one for each valley $\pm K$.
We focus on only the $-K$ valley,
\begin{eqnarray}\label{eq_cm}
H_{- K}^{b,t}&=&-\frac{(\bm{k} - \bm{\kappa}^{(\prime)})^2}{2m_{b,t}}+V_{b,t}(\bm r),\\\label{eq_v}
V_{b,t}(\bm r)&=&2V_{b,t}\sum_{i=1,3,5}\cos(\bm g_{i}\cdot\bm r+\phi_{b,t}),
\end{eqnarray}
where $\bm{k}=-i\bm{\nabla}$,
 and $m_{b,t}>0$ is the effective mass in MoTe$_2$ (bottom layer) and WSe$_2$ (top layer). The potential term is expressed as the first-order harmonics with  moir\'e wave vectors $\bm g_j=\frac{4\pi}{\sqrt{3}a_M}(-\sin\frac{\pi (j-1)}{3},\cos\frac{\pi (j-1)}{3})$ \cite{Wu2018}.
Due to the BZ folding, take the minimum of the dispersion, $\bm{\kappa}^{(\prime)}$, to be $\bm{\kappa}=\frac{4\pi}{3 a_M}(-\frac{1}{2},\frac{\sqrt{3}}{2})$ for the bottom layer, and $\bm{\kappa}^\prime=\frac{4\pi}{3 a_M}(\frac{1}{2},\frac{\sqrt{3}}{2})$ for the top layer.


The continuum Hamiltonian involving both layers is given by
\begin{equation}
H_{-K} = \begin{pmatrix}
H^{b}_{-K} & W(\bm{r}) \\
W^\dagger(\bm{r}) & H^{t}_{-K} - \Delta E_g
\end{pmatrix}
\end{equation}
where $m_{t,b}$ are effective masses, $V_{b,t}(\bm{r})$ is an effective potential arising due to corrugation effects, $W(\bm{r})$ is an interlayer tunneling term,
and $\Delta E_g$ is the energy energy between the two layers which can be tuned experimentally by the displacement field.
The continuum model has three-fold counter-clockwise rotation symmetry $C_{3z}$ about the origin, chosen to be between the metal atoms of the two layers at an MM stacking region.

The interlayer tunneling term is
\begin{equation}
W^\dagger(\bm{r}) = W(1 + \omega^{\nu} e^{i \bm{g}_2 \cdot \bm{r}} + \omega^{2\nu} e^{i \bm{g}_3\cdot\bm{r}})
\end{equation}
where $\omega=e^{2\pi i / 3}$, and $C_{3z}$ symmetry fixes $\nu$ depending on the difference in $C_{3z}$ eigenvalue of the $-K_{b,t}$ states in the two layers.
Specifically, if the state at $-K_{b(t)}$ on the bottom (top) layer, $\ket{-K_{b (t)}}$, has $C_{3z}$ eigenvalue $\omega^{n_{b (t)}}$, then $\nu=n_b-n_t$ mod $3$.
The eigenvalues $\omega^{n_{b,t}}$ can be determined from monolayer DFT calculations by computing the $C_{3z}$ eigenvalues of the valence band top at $-K$. We find $n_b=-\frac{1}{2}$ and $n_t=\frac{1}{2}$.

Since the $-K$ valley MoTe$_2$ band is mostly spin-$\downarrow$, and WSe$_2$ band mostly spin-$\uparrow$,
these $C_{3z}$ eigenvalues can be decomposed as $n_b = -1 + \frac{1}{2}$ and $n_t=1-\frac{1}{2}$, where the integer part is due to the orbital wavefunction and the half-integer part is due to spin.

Again, we note that if the spin-$U(1)$ symmetry was exact and $s_z$ a good quantum number, interlayer tunneling would be forbidden.
However, as observed in the DFT band structure, there is indeed interlayer hybridization between bands of different spin.
Thus, the interlayer tunneling is non-zero, albeit weak, and arises due to the fact that the bands are not fully spin polarized.
The continuum model is agnostic of the microscopic origin of the interlayer tunneling,
and $W(\bm{r})$ is simply the first harmonic approximation of a generic complex-valued tunneling term that is allowed by $C_{3z}$ symmetry.

This continuum model differs from the previously studied continuum model for TMD heterobilayers~\cite{Wu2018}
as it takes into account both layers that are coupled through interlayer tunneling, rather than just keeping a single layer, which is only valid in the limit of large $\Delta E_g$.
At the displacement fields at which the band inversion takes place, both layers are important.
The continuum model also differs from that of AA stacked TMD homobilayers~\cite{Wu2019} with a displacement field,
for which the interlayer tunneling term always has $\nu=0$ due to the alignment of spins on the two layers.
Depending on $\nu$, the tunneling $|W(\bm{r})|$ will be peaked at one of either MM, MX, or XX regions and zero at the other two, due to the difference in $C_{3z}$ eigenvalue.

The BZ folding (Fig~\ref{fig:mono}) is such that the $-K_b$ MoTe$_2$ band maximum is at $\kappa^\prime$ in the mBZ, with narrow gaps opened at $\gamma$ and $\kappa$ at energy $E\approx -\frac{|\bm{\kappa}|^2}{2_{m_b}}$.
Meanwhile, the $-K_t$ WSe$_2$ band maximum is located at $\kappa$ at energy $E\approx -\Delta E_g$.
By tuning displacement field to $\Delta E_g\sim \frac{|\bm{\kappa}|^2}{2 m_b}$, the WSe$_2$ valence band top can cause a band inversion with the MoTe$_2$ bands at $\kappa$.

Let us focus on the bands at $\kappa$ relevant to the band inversion.
When $V_b=W=0$, there are three exactly degenerate MoTe$_2$ bands at $\kappa$, given in the continuum model by plane waves $\{\ket{-\bm{K}_b + \bm{\delta}_j}\}_{j\in\{1,2,3\}}$, where $\bm{\delta}_j$ are counter-clockwise $2\pi(j-1)$ rotations of  $\bm{\delta}_1=\frac{4\pi}{3 a_M}(1,0)$.
The $C_{3z}$ symmetry acts as
$\ket{-\bm{K}_b+\bm{\delta}_j} \rightarrow \omega^{n_b} \ket{-\bm{K}_b+\bm{\delta}_{j+1}}$.
For non-zero $V_b$, the degeneracy is split and the resulting eigenstates (restricted to this three-state subspace) are $C_{3z}$ eigenstates
\begin{equation}
\ket{n} = \frac{1}{\sqrt{3}}\sum_{j=1,2,3} \omega^{j n} \ket{-\bm{K}_b + \bm{\delta}_j} \\
\end{equation}
with energy
\begin{equation}
E_n=-\frac{|\bm{\kappa}|^2}{2m_b}+ 2V_b \cos(2\pi n / 3 + \phi_b)
\end{equation}
 and $C_{3z}$ eigenvalue $\omega^{n_b - n}$.
These eigenstates also have characteristic spatial dependence.
Let $\bm{R}_m=\frac{m a_M}{\sqrt{3}}(\frac{\sqrt{3}}{2},\frac{1}{2})$, for $m\in\{0,1,2\}$, be the high symmetry positions MM ($\bm{R}_0$), MX ($\bm{R}_1$), and XX ($\bm{R}_2$).
Then, $|{\braket{\bm{R}_m|n}}| \propto |1+\omega^{m-n} + \omega^{n-m}|$ is maximized at $\bm{R}_{m=n}$ and zero at $\bm{R}_{m\neq n}$.
Thus, the state $\{\ket{n}\}$ is peaked at the high symmetry position $\bm{R}_n$, and their energy splitting is also proportional to the potential $V_b(\bm{R}_n)$.

We can obtain the relevant continuum model parameters by fitting to the band structure at displacement field $-0.3$V/nm.
Since the essential physics involves only the band maximum of WSe$_2$, the potential $V_t(\bm{r})$ is unimportant and we simply set $V_t=0$.
From fitting to the energies, and using the long wavelength features of the DFT majority-spin wavefunctions at $\kappa$, we find the following parameters:
($V_b$,$\phi_b$,$W$)=($4.1$meV,$14^\circ$,$1.3$meV),
($m_b$,$m_t$)=$(0.65 m_0$,$0.35 m_0$) where $m_0$ is the electron mass (using lattice constants $a_b=3.575$\AA{} and $a_t=3.32$\AA{}),
and $\Delta E_g = 36.8$meV.
The fitted effective mass is very close to the monolayer value reported in previous experimental and theoretical study \cite{fallahazad2016shubnikov,rasmussen2015computational}.

\begin{figure}[t]
\includegraphics[width=\columnwidth]{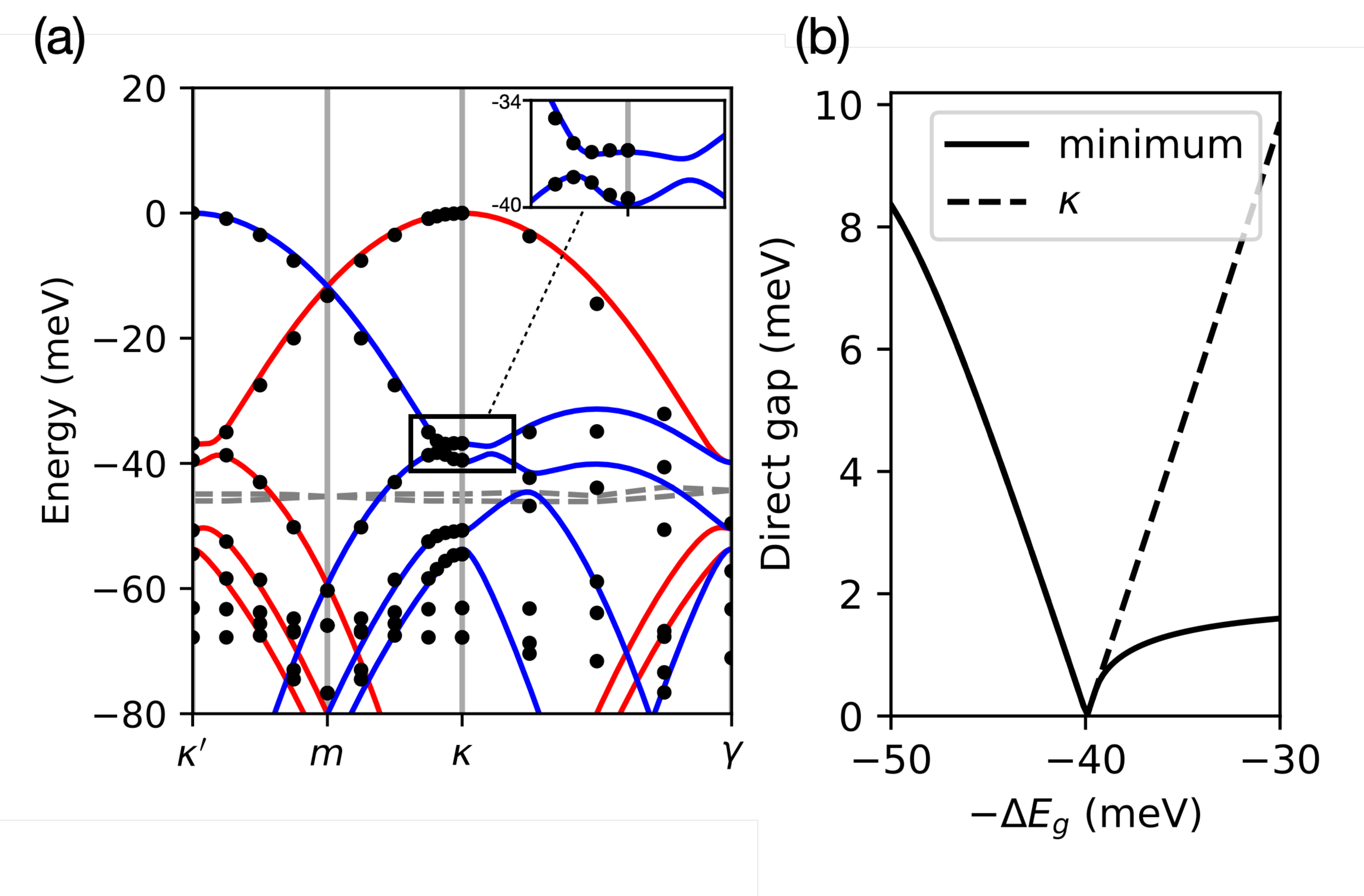}
\caption{
(a) Band structure of the continuum model with parameters mentioned in the main text.  Blue lines indicate $-K$ valley bands and red lines indicate $+K$ valley bands.
The black dots represent the DFT bands computed for the heterobilayer at displacement field $-0.3$V/nm (dashed lines are states originating from the Gamma pocket).
Inset shows the avoided crossing due to interlayer hybridization near $\kappa$.
(b) The direct gap between the first two $-K$ valley bands at $\bm{k}=\bm{\kappa}$, and minimized over $\bm{k}$, as a function of $\Delta E_g$.
The first band has non-trivial Chern number for $\Delta E_g < 40$meV.
}\label{fig:continuum}
\end{figure}

The resulting continuum model bands are shown in Fig~\ref{fig:continuum}a, in comparison with DFT bands.
The continuum model agrees very well with DFT along the $m$--$\kappa$ path, but differs slightly near $\gamma$,
as a result of neglected time reversal symmetry breaking terms within one valley which make $\gamma$ and $\kappa$ nonequivalent.
This effect, present even in the corrugated monolayer, arises from spatially varying strain that produces valley-contrasting
pseudomagnetic field \cite{xie2021theory,zhai2020theory},
but does not play an important role in the band inversion in MoTe$_2$/WSe$_2$.
In Fig~\ref{fig:continuum}b, we show the direct gap between the first two bands at $\kappa$, as well as the minimum over all $\bm{k}$, as a function of  $\Delta E_g$.
The gap closes at $\kappa$ at $\Delta E_g\approx 40$meV, and the first band remains topological below it.
We compute the Chern number from the continuum model after inversion to be $\mathcal{C}_{-K}=1$.
The $+K$ valley band will have opposite Chern number due to time reversal symmetry.

Importantly, the physics of topological band inversion in TMD bilayers is essentially
dictated by the $C_{3z}$ symmetry eigenvalues of the top MoTe$_2$ and WSe$_2$ bands at $\kappa$.
Provided that $C_{3z}$ eigenvalues of the two are different, decreasing the layer bias potential beyond $\Delta E_g \lesssim \frac{|\bm{\kappa}|^2}{2 m_b}$ will close the gap and drive a  band inversion.
Prior to band inversion, the first $-K$ valley band at $\kappa$ has $C_{3z}$ eigenvalue $\omega^{n_b-n_{0}}$ at $\kappa$, where $n_0=-\lfloor\frac{\phi_b+\pi/3}{2\pi/3}\rfloor$mod $3$ is determined by the location of the maximum of $V_b(\bm{r})$ ($n_{0}=0$ for MoTe$_2$/WSe$_2$).
After the band inversion, the state at $\kappa$ of the first band will instead have $C_{3z}$ eigenvalue $\omega^{n_t}$, and consist mainly of the state $\ket{-K_t}$ of the top layer.
The resulting band after inversion has $C_{3z}$ eigenvalues $\omega^{n_b}$ at $\kappa^\prime$, $\omega^{n_b+n_0}$ at $\gamma$, and $\omega^{n_t}$ at $\kappa$.
The interlayer tunneling $W\neq 0$ is crucial in opening a direct gap in the inverted regime.
As long as $n_t\neq n_b-n_0\mod 3$, this eigenvalue structure cannot be captured by a single band tight-binding model with any choice of Wannier centers, and will have non-trivial Chern number.
Note that here, $n_t\neq n_b$ due to the different spin of $-K$ valley monolayer bands in the AB stacked case.
In the AA stacked case, we would have had $n_t=n_b$, and the resulting bands would not be topological for $n_0=0$.

We note that the relative $C_{3z}$ eigenvalues at $\kappa$, $\kappa^\prime$, and $\gamma$ dictate the position of the Wannier centers for the first moir\'e bands from each layer which participate in band inversion.
For the first $-K$ valley MoTe$_2$ moir\'e band, all $C_{3z}$ eigenvalues are identical, $\omega^{-\frac{1}{2}}$, indicating that the Wannier centers should be at the MM regions.
For WSe$_2$, we calculate the $C_{3z}$ eigenvalues directly from DFT to be $\omega^{\frac{1}{2}}$,$\omega^{-\frac{1}{2}}$, and $\omega^{\frac{3}{2}}$, at $\kappa$,$\kappa^\prime$, and $\gamma$.
This eigenvalue structure corresponds to Wannier functions centered at the XX region, consistent with the peak of the DFT wavefunctions at the XX region.
A tight binding description of the first bands from each layer is therefore a honeycomb lattice with sites centered at the MM and XX regions of the moir\'e superlattice.

\textbf{Discussion}
The low energy theory near topological band inversion in AB-stacked TMD bilayers is described by a massive Dirac Hamiltonian
\begin{equation}
H_{{D}} = v (q_x \sigma_x + q_y \tau_z \sigma_y) + m \sigma_z
\end{equation}
where $\sigma_\alpha$ and $\tau_\alpha$ are layer and valley Pauli matrices respectively, and $\bm{q} \equiv \bm{k} - \tau_z \bm{\kappa}$.
The Dirac mass determines the band gap between first and second moir\'e bands at $\kappa, \kappa'$ and is controlled by the displacement field. The sign reversal of Dirac mass $m$ leads to a change of valley Chern number from $C=0$ to $C=1$.
The displacement field induced band inversion in MoTe$_2$/WSe$_2$
is reminiscent of the case of InAs/GaSb quantum wells \cite{liu2008quantum,du2015robust}.
Due to spin-valley locking, the complete filling of Chern bands in both valleys, corresponding to 2 holes per moir\'e unit cell,
gives rise to a quantum spin Hall insulator.

While our calculation focused on MoTe$_2$/WSe$_2$,
the general mechanism for band inversion we uncovered in AB-stacked TMD bilayers
is quite robust, and can be applied to other material combinations as well.
In such bilayer systems with a tunable band offset,
the formation of valley-contrasting Chern bands depends only on the difference of $C_{3z}$ eigenvalues
between the bands at the band inversion point.

Including the effect of Coulomb interaction in these valley-contrasting Chern bands
may lead to a number of interesting electronic phases, such as the recently observed QAH insulator at half band filling
in AB-stacked MoTe$_2$/WSe$_2$ \cite{li2021quantum}.
This phase may occur as a result of interaction-induced valley polarization.
In this scenario, holes would fully fill a Chern band from one of the two valleys $K$ or $-K$,
realizing a QAH state at filling of one hole per moir\'e unit cell.
Our comprehensive DFT calculation and theoretical analysis show that
the envelope wavefunctions of the two dispersive bands form a honeycomb lattice structure.
Therefore a theoretical framework for studying interaction effects in MoTe$_2$/WSe$_2$ is
the Kane-Mele-Hubbard model. As shown in our recent work~\cite{devakul2021magic}, this model exhibits a variety of insulating phases at half band filling, including the QAH and Mott insulating phases. A detailed phase diagram for MoTe$_2$/WSe$_2$ will be presented in our follow-up work.
Finally, we remark that specific details such as the negative overlap, $\Gamma$ pocket bands, and interaction-induced band renormalization, will depend on the specific materials, moir\'e wavelength, and the choice of van der Walls density functionals; however, the displacement field induced band inversion and band topology is generic and depends only on the $C_{3z}$ eigenvalues and non-zero interlayer tunneling.




\textbf{Method}
We perform the density functional calculations using generalized gradient approximation \cite{perdew1996generalized} with SCAN+rVV10 van der Waals density functional \cite{peng2016versatile}, as implemented in the Vienna Ab initio Simulation Package \cite{kresse1996efficiency}. Pseudopotentials are used to describe the electron-ion interactions.  We first construct AB stacked MoTe$_2$/WSe$_2$ heterobilayer with vacuum spacing larger than 20 A to avoid artificial interaction between the periodic images along the $z$ direction. Dipole correction is added to the local potential in order to correct the errors introduced by the periodic boundary conditions in out of plane direction. The structure relaxation is performed with force on each atom less than 0.01 eV/A. We use Gamma-point sampling for structure relaxation and self-consistent calculation, due to the large moire unit cell.

\section*{Acknowledgment}
We thank Kin Fai Mak, Jie Shan, Tingxin Li and Shengwei Jiang for collaboration on a parallel experimental work. YZ acknowledges Yan Sun and Claudia Felser for providing the computational resources at MPCDF Raven cluster.
This work is primarily supported by DOE Office of Basic Energy Sciences, Division of Materials Sciences and Engineering under Award  DE-SC0018945 (theoretical modeling) and DE-SC0020149 (band structure calculation).
LF is partly supported by  Simons Investigator award from the Simons Foundation
and the David and Lucile Packard Foundation.

\bibliography{ref}

\end{document}